\documentclass[%
 reprint,
 amsmath,amssymb,
 aps,
 pra,
floatfix,
superscriptaddress]{revtex4-2}

\usepackage{graphicx}
\usepackage{dcolumn}
\usepackage{bm}
\usepackage{booktabs}
\usepackage[english]{babel}
\usepackage{hyperref}
\usepackage[capitalise]{cleveref}
\usepackage{lipsum}
\usepackage{glossaries}
\usepackage{tikz}
\usepackage{pgfplots}
\pgfplotsset{compat=1.18}
\usetikzlibrary{arrows.meta, shapes.misc}
\usetikzlibrary{fit,backgrounds,positioning}
\usepackage{siunitx} 
\usepackage{braket}
\usepackage{bm}
\glsdisablehyper

\newacronym{ie}{IE}{information engine}
\newacronym{tls}{TLS}{two-level system}
\newacronym{qho}{QHO}{quantum harmonic oscillator}
\newacronym{cop}{COP}{coefficient of performance}

\newcommand{\tr}[2][]{\operatorname{tr}_{\text{#1}}\left\{#2\right\}}
\newcommand{\dm}[1][]{\hat{\rho}_{\text{#1}}}
\newcommand{\Ham}{\hat{H}}
\newcommand{\HamS}{\hat{H}_{\text{S}}}
\newcommand{\HamSM}{\hat{H}_{\text{S/M}}}
\newcommand{\HamM}{\hat{H}_{\text{M}}}
\newcommand{\HamI}{\hat{V}_{\text{I}}}
\newcommand{\Wmeas}{W_{\text{meas}}}
\newcommand{\Wext}{W_{\text{ext}}}

\newcommand{\tm}{t_{\text{m}}}
\newcommand{\kb}{k_{\text{B}}}
\newcommand{\Temp}[1][]{T_{\text{#1}}}

\newcommand{\geff}{g_\text{eff}}
\DeclareMathOperator{\diag}{diag}
\DeclareMathOperator{\rank}{rank}

\definecolor{revcolor}{rgb}{0.78,0.10,0.10}

\newif\ifulemloaded
\IfFileExists{ulem.sty}{\usepackage[normalem]{ulem}\ulemloadedtrue}{\ulemloadedfalse}
\ifulemloaded
  
\else
  
\fi

\begin{document}

\title{Pareto-optimal work extraction and the thermodynamic cost of precision in quantum information engines}

\author{Jonas Berx}%
\email{jonas.berx@nbi.ku.dk}
\affiliation{%
 Niels Bohr International Academy, Niels Bohr Institute, University of Copenhagen, Blegdamsvej 17,
2100 Copenhagen, Denmark
}%

\author{Ted Olander}
\affiliation{Department of Microtechnology and Nanoscience (MC2), Chalmers University of Technology, S-412 96 Göteborg, Sweden}

\author{Henning Kirchberg}
\email[]{henning.kirchberg@chalmers.se}
\affiliation{Department of Microtechnology and Nanoscience (MC2), Chalmers University of Technology, S-412 96 Göteborg, Sweden}

\date{\today}

\begin{abstract}
We study a finite-time quantum information engine in which a two-level system is measured by a quantum harmonic oscillator acting as a meter and where useful work is extracted conditionally on the measurement outcome. Using multi-objective optimisation, we find a Pareto-optimal trade-off between extractable work and its fluctuations and show that reducing fluctuations entails higher thermodynamic costs: greater information consumption, more engine cycles, longer operation time, and reduced average work output. In the limit of a highly accurate meter, we obtain the work distribution, its moments, and the Pareto front analytically. In this regime, the work statistics of the engine reduce to those of a qubit in contact with a single thermal bath. We further analyse the associated information flows by examining the mutual information and Fisher information, and show that the Pareto-optimal engine designs lie very close to local maxima of the latter with respect to the operation time of the device. Our results provide a compact description of the trade-offs between work, its fluctuations, and thermodynamic costs in quantum information engines. 
\end{abstract}

\maketitle


\section{\label{sec:introduction}Introduction}

Converting the heat flowing between hot and cold baths into useful energy is the paradigmatic operation of a heat engine~\cite{Scovil1959,Alicki_1979,Kosloff1984}, whose energy conversion performance is fundamentally constrained by the second law of thermodynamics. In particular, this constraint forbids work extraction from a single thermal bath~\cite{Jarzynski1997,Campisi2009}. 

Advances in high-precision measurement and control with high temporal resolution have, however, opened a complementary route to energy conversion by exploiting information gained through measurement~\cite{Szilard1929,Maruyama2009Jan,junior2025}.
Measurement and feedback can be consistently incorporated into the second law~\cite{SagawaUeda2008}, demonstrating that information can serve as a genuine thermodynamic resource. As a result, information-powered engines can transduce heat into useful energy from only a single heat bath, a principle that has now been realised in both classical and quantum experiments~\cite{Toyabe2010,Koski2014,Monsel2025Jan,Kirchberg2025}.

Despite this progress, fluctuations become non-negligible when measurement-controlled energy transduction is implemented in small-scale devices. At this small scale maximising the \textit{average} extracted useful energy alone is no longer a sufficient performance criterion~\cite{Campisi2009,Miller2019,Kirchberg2022Noise,Tesser2024}. Instead, useful energy must be delivered \textit{reliably}, i.e., with minimal cycle-to-cycle variability. This requirement imposes stringent constraints on the precision of the underlying stochastic processes~\cite{Jarzynski1997,Crooks1999,Hanggi2015}, elevating fluctuations from a secondary effect to a central design consideration of the energy transduction process. The issue is particularly acute for measurement-based engines, where work extraction is conditioned on stochastic measurement outcomes, yielding a finite energy output only in a subset of “successful” cycles. The key question is therefore not only how much work is extracted on average, but whether information-driven extraction can be made \emph{reliable}.

Quantum implementations of information engines have recently become experimentally accessible~\cite{Spiecker2023Szilard,Zhang2025}. In particular, qubit-waveguide platforms demonstrate how information about a qubit can be converted into useful energy via stimulated emission within a finite operation time~\cite{Cottet2017,Masuyama2018,Dassonneville_2026}. In these settings, fluctuations in energy transduction and output directly translate into operational errors, limiting performance and scalability in size and operation time.

Designing quantum engines that exploit information for energy extraction therefore requires navigating a fundamental trade-off between \textit{output} and \textit{precision}. While maximising extracted work is desirable, it generally amplifies fluctuations and compromises reliability. Conversely, suppressing fluctuations and improving reliability typically incurs a higher thermodynamic cost.
However, a comprehensive understanding of the \textit{global thermodynamic costs} associated with measurement and the subsequent information acquisition remains elusive. Such a perspective must extend beyond energetic expenditures and dissipation to include temporal resources, such as operation time or the number of operation cycles, as well as information itself as a consumable time-dependent resource. 

In this work, we present a systematic exploration of this global framework by investigating a quantum information engine inspired by a qubit-waveguide setup that has been experimentally realised~\cite{Masuyama2018} and that we previously studied theoretically~\cite{Hagman2025} regarding the \textit{average} extractable work. Here, we optimise the engine to maximise the extractable work while minimising work fluctuations, thereby revealing the trade-offs between different thermodynamic costs. This approach reveals the structure of the work-fluctuation trade-off and identifies operational regimes that remain hidden in single-objective analyses. 

To characterise these trade-offs and navigate the high-dimensional control landscape of the engine, we implement a multi-objective optimisation using a genetic algorithm~\cite{Deb2008}, leading to \textit{Pareto fronts}. This multi-objective optimisation framework is widely used in engineering and economics and has only recently been applied to nonequilibrium thermodynamics~\cite{Berx_2024_2,Forao2025,Berx_2025} and quantum devices~\cite{Hagman2025,Erdman2023}. We show that Pareto optimality among the different \textit{thermodynamic costs} emerges naturally as the key principle to design reliable measurement-driven engines with high performance.

This paper is organised as follows. In Secs.~\ref{sec:method} and~\ref{sec:method_example}, we briefly introduce the general setup of the quantum information engine and illustrate it for the case of a two-level system coupled to a quantum harmonic oscillator. In Sec.~\ref{sec:results}, we introduce the performance measures, analyse their Pareto optimisation and comment on experimental finite-cycle limitations. In Sec.~\ref{sec:information}, we examine the associated information flows and sensitivity. Finally, Sec.~\ref{sec:conclusion} presents our conclusions and an outlook on future directions.

\begin{figure}[htp]
    \centering
    \includegraphics[width=\linewidth]{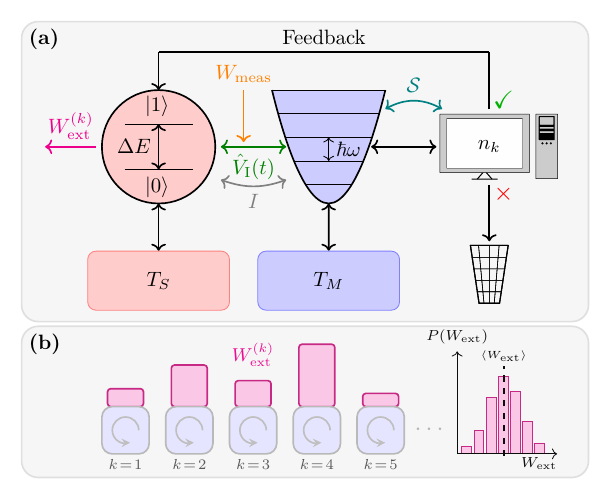}
    \caption{{\bf(a)} Schematic of a single information-engine cycle, showing a concrete  model in which the \textit{system} is a two-level system (\gls{tls}) and the \textit{meter} is a quantum harmonic oscillator (\gls{qho}). Coupling them via the time-dependent interaction $\hat{V}_{\rm I}$ costs the measurement work $W_\mathrm{meas}$ and goes along with information transfer $I$. In the $k$-th cycle, the \gls{qho} is projectively measured in its energy eigenstate $n_k$ with a classical measurement device with a corresponding entropy flow $\mathcal{S}$. Via feedback a stimulated emission is initiated, leading to the extracted work $\Wext^{(k)}$, which can be zero for unfavourable measurement outcomes $n_k$. After each cycle the \gls{tls} and the \gls{qho} are coupled to thermal baths at temperatures $\Temp[S]$ and $\Temp[M]$, respectively. {\bf(b)} Extracted work statistics. Each engine cycle $k$ (blue boxes) extracts work $\Wext^{(k)}$ (magenta boxes), leading to a work distribution $P(\Wext)$.} 
    \label{fig:ie_model}
\end{figure}

\section{\label{sec:method} Information-engine principles}

Our quantum information engine consists of a system (S), from which work is extracted, and a meter (M) that obtains information about the system. The system and the meter are coupled for a certain time $\tm$, become correlated, and, due to this correlation, information about the system can be obtained through projective measurement of the meter. This information can subsequently be exploited to extract work from the system. The system-meter setup is captured by the Hamiltonian
\begin{equation}\label{eq:ie_model_general}
    \Ham(t) = \HamS + \HamM + \HamI(t)
\end{equation}
where $\HamS$ is the Hamiltonian of the system, $\HamM$ is the Hamiltonian of the meter, and $\HamI(t)$ describes their time-dependent interaction. We take the interaction $\HamI(t)$ to be finite in a time interval $t\in(0,t_\mathrm{m})$ and zero otherwise. 

While system and meter are not coupled, both interact with their respective heat baths at inverse temperatures $\beta_S = 1/\kb\Temp[S]$ and $\beta_M = 1/\kb\Temp[M]$. We refer to these two baths as the system and meter baths, respectively.

\textbf{Engine cycle--} The engine cycle consists of the following steps.
\paragraph{Initialization:}
The system and meter are initially uncoupled and in thermal states
\begin{align}
\label{eq:sys_in}
\hat{\rho}_\mathrm{S/M}(0) = \frac{\exp (-\beta_{S/M}\HamSM)}{\tr{\exp (-\beta_{S/M}\HamSM)}}\,, 
\end{align}
such that the total initial state is $\dm(0) = \dm[S](0)\otimes\dm[M](0)$.

\paragraph{Unitary evolution:} 
The system and meter are decoupled from their respective baths and subsequently coupled to each other during $0 < t < \tm$, with $\tm$ the measurement time. Importantly, since $[\HamI(t), \HamM] \neq 0$, system and meter become correlated during this period. Assuming instant switching of the coupling Hamiltonian at times $t=0$ (``on") and at $t=\tm$ (``off"), the measurement work, namely the work spent to correlate the system and meter states, is given by
\begin{equation}\label{eq:measurement_work_definition}
    \Wmeas(\tm) \equiv \tr{ [\HamS+\HamM]\big(\dm(\tm)-\dm(0) \big) },
\end{equation}
where $\dm(t) = \hat{U}(t) \dm(0)\hat{U}^\dagger(t)$ with $\hat{U}(t) = \mathcal{T}\exp{\left[-\frac{i}{\hbar}\int_0^tdt' \Ham(t')\right]}$, and $\mathcal{T}$ being the time-ordering operator.

\paragraph{Projective measurement \& information acquisition:}
At $t=\tm$ the coupling between system and meter is turned off.
The state of the meter is then determined by a projective measurement on the eigenstates $\ket{n}$ of $\HamM$ by a classical meter~\cite{Taranto2023}. This quantum-to-classical transition is the Heisenberg cut~\cite{Heisenberg1949,VonNeumannBook,Atmanspacher01091997}.  After the projective measurement, the density matrix conditioned on the measurement outcome $n$ is
\begin{equation}\label{eq:conditional_density_matrix}
    \hat{\rho}_{\rm S}(\tm|n) = \frac{\braket{n|\dm(\tm)|n}}{\tr[S]{\braket{n|\dm(\tm)|n}}} \equiv \frac{\braket{n|\dm(\tm)|n}}{P(n,\tm)},
\end{equation}
which depends on the measurement outcome $n$.
The conditional probability of finding the system in state $\ket{i}$ given the measurement outcome $n$, is given by the diagonal elements of $\hat{\rho}(\tm|n)$, i.e., $P(i|n,\tm) = \braket{i|\dm(\tm|n)|i}$. Similarly, the joint probability of the meter being in state $\ket{n}$ and the system being in state $\ket{i}$ at time $\tm$ is given by $P(i,n,\tm) = \braket{i| \braket{n|\dm(\tm)|n}| i}$. 

Starting from Eq.~\eqref{eq:conditional_density_matrix}, the conditional entropy $S_n(\tm)$ given the measurement outcome $n$ reads $S_n(\tm) = -\kb \tr[S]{\hat{\rho}(\tm|n) \ln \hat{\rho}(\tm|n)}$. Averaging over measurement outcomes, the entropy $S$ then becomes $S(\tm) = \sum_n P(n,\tm) S_n(\tm)$, with $P(n,\tm)$ as defined in~\cref{eq:conditional_density_matrix}. We quantify the information gain during the measurement process by the entropy difference~\cite{Kirchberg2025}: 

\begin{equation}
\label{eq:InformationGain}
 I \equiv  S(0) - S(\tm) .  
\end{equation}
\paragraph{Work extraction:}
Given the measurement outcome $n$, 
the conditional extractable work is given by \emph{ergotropy}: the maximal work that can be extracted under unitary transformation $\hat{U}_n$ conditioned on $n$~\cite{Allahverdyan2004MaximalSystems,Francica2017DaemonicCorrelations}, i.e.,

\begin{align}
\label{eq:work_extraction}
    \Wext (\tm|n) = & \tr{\dm[S](\tm|n)\HamS} \\ \notag &- \min_{\hat{U}_n} \tr{ \hat{U}_n\dm[S](\tm|n)\hat{U}_n^\dagger\HamS }, 
\end{align}
which is positive or zero depending on the outcome of the measurement $n$. 

Equation~\eqref{eq:work_extraction} allows one to determine all work moments by computing the ensemble average over many engine cycles. For instance, the $j$th work moment is given by
\begin{equation}
\label{eq:work_moments}
    \langle\Wext^{j} (\tm)\rangle = \sum_n P(n,\tm) \left[\Wext(\tm|n)\right]^j.
\end{equation}

\paragraph{Resetting:}
The system and the meter are coupled to their respective baths, allowing both to rethermalise, closing the cycle. In principle, we must include the Landauer erasure cost~\cite{Landauer1961} of resetting the classical measurement device (see~\cref{fig:ie_model}), which displays and stores the measurement outcomes. However, for a classical readout performed at arbitrarily low temperature, this erasure work is negligible~\cite{Elouard2017,Jordan2020,Taranto2023,Hagman2025}.

\section{\label{sec:method_example} Two-level system measured by a harmonic oscillator}
To illustrate the general model introduced in Sec.~\ref{sec:method}, we consider a quantum two-level system (\gls{tls}) with energy-level spacing $\Delta E$ as the system S and a quantum harmonic oscillator (\gls{qho}) of frequency $\omega$ and mass $\mathcal{M}$ as the meter M, as shown in~\cref{fig:ie_model}(a). The Hamiltonians of the system and the meter are given by
\begin{equation}
    \HamS = \Delta E \ket{1}\bra{1}, \qquad \HamM = \frac{\mathcal{M}\omega^2}{2}\hat{x}^2 + \frac{1}{2\mathcal{M}}\hat{p}^2\,,
\end{equation}
where the energy of the $\ket{0}$ state is set to zero. The interaction Hamiltonian $\HamI(t)$ is switched on for the measurement time only, i.e.,
\begin{equation}\label{eq:interaction_hamiltonian}
    \HamI(t) = \begin{cases}
        g \ket{1}\bra{1}\otimes\hat{p},\quad t\in (0,\tm) \\
        0, \quad t \notin (0,\tm).
    \end{cases}
\end{equation}
This form of coupling is motivated by qubit-waveguide architectures where the waveguide hosts a selected photonic mode that couples to the qubit, with the interaction strength being dynamically controllable through an external magnetic flux~\cite{Janzen2023}.

The initial state of the combined system and meter is 
\begin{align}
\label{eq:InitialStateExample}
    \hat{\rho} (0) &= \dm[S](0)\otimes\dm[M](0) \\ \notag   
    &= \big(a\ket{0}\bra{0} + b\ket{1}\bra{1}\big) \otimes \sum_{n=0}^\infty \frac{1}{Z_M}e^{-\beta_M\HamM}\ket{n}\bra{n},
\end{align}
where $a=(1+e^{-\beta_\text{S}\Delta E})^{-1}$ and $b= 1-a$ are the initial populations of the ground and excited states of the \gls{tls}, respectively. The initial meter state in Eq.\ \eqref{eq:InitialStateExample} is thermal, defined by the partition function $Z_\text{M} = \tr{e^{-\beta_\text{M} \HamM}}$. 

During the subsequent coupled evolution the system and the meter evolve unitarily under the Hamiltonian $\Ham(t)=\HamS+\HamM+\HamI(t)$, resulting in a correlated state. The state of the full system at time $\tm$ is thus given by
\begin{eqnarray}
\label{eq:Unitary}
    \hat{\rho}(\tm) = \hat{U}(\tm)\dm(0)\hat{U}^\dagger(\tm).
\end{eqnarray}
The unitary state evolution in Eq.~\eqref{eq:Unitary}, in which the effective coupling to the respective heat baths is neglected, is well justified for qubit-waveguide setups. In such systems, the relevant measurement and control operations occur on timescales much shorter than the relaxation and decoherence times; see, for example, Ref.~\cite{Yang2025}.

By evaluating the measurement cost,~\cref{eq:measurement_work_definition}, for the case of a sudden switch of the system-meter interaction~\eqref{eq:interaction_hamiltonian}, we find 
\begin{equation}
    \label{eq:measurement_work_specific}
    \begin{split}
        \Wmeas(\tm) &= bg^2_{\text{eff}}\Big(1-\cos(\omega \tm) \Big), 
    \end{split}
\end{equation}
where $g_{\text{eff}}^2 = g^2\mathcal{M}$ is the effective coupling strength. Performing the projective measurement at $t=t_m$, the joint probabilities $P(i,n,\tm)$ of finding the \gls{tls} in state $\ket{i}$ and the \gls{qho} in state $\ket{n}$ are given by
\begin{align} \label{eq:probability0}
    P(0,n,\tm) &= a \left(1-e^{-\beta_\text{M}\hbar\omega}\right)e^{-\beta_\text{M}\hbar\omega n} \\ \label{eq:probability1}
    P(1,n,\tm) &= \\ b\sum_m P_m &\left|\left(\frac{m!}{n!}\right)^{1/2}\alpha^{n-m}e^{-|\alpha|^2/2}L_m^{(n-m)}\left(|\alpha|^2\right)\right|^2 \notag
\end{align}
where $L_m^{(n-m)}(x)$ are the generalised Laguerre polynomials~\cite{Magnus2013} and 
\begin{equation}
 \alpha=\frac{\geff}{\sqrt{2\hbar \omega}}\Big(\sin(\omega \tm)-i\Big(\cos(\omega \tm) -1\Big)\Big)
\end{equation}
is the coherent-state displacement in the position-momentum phase space. The corresponding conditional probabilities are then given by $P(i|n,\tm) = P(i,n,\tm)/\sum_iP(i,n,\tm)$.

Since we describe a non-demolishing measurement in this example, $[\hat{H}_{\rm S},\hat{V}_{{\rm I}}]=0$, the work extracted from the information engine is bounded by $\Wext(\tm) \leq T_{\rm S} I(\tm)$ (see~\cite{Hagman2025}).

Given a measurement outcome $n$ of the meter, we consider the work extraction by ergotropy, Eq.~\eqref{eq:work_extraction}.
For a \gls{tls} this corresponds to population inversion, such that the extracted energy in this case is
\begin{equation}
\label{eq:work_ext_TLS}
    \Wext(\tm|n) = \Delta E\,
    \Pi_n(\tm)\,    \Theta \big(\Pi_n(\tm)\big),
\end{equation}
where $\Theta(x)$ is the Heaviside function, and $\Pi_n(\tm) \equiv P(1|n,\tm) -P(0|n,\tm)$ is the \textit{conditional polarisation} of the TLS. Averaging over several engine cycles, the ergotropy reads
\begin{equation}
\label{eq:work_extracted_specific}
    \begin{split}
        \langle \Wext(\tm) \rangle &= \sum_nP(n,\tm) \Wext(\tm|n) \\
        &= \Delta E \sum_nP(n,\tm)
        \Pi_n(\tm)\,    \Theta \big(\Pi_n(\tm)\big)
    \end{split}
\end{equation}

Similarly, using~\cref{eq:work_moments}, the second moment reads
\begin{align}
\label{ergotropysecond}
    \langle \Wext^2 (\tm) \rangle &= \Delta E^2 \sum_{n}^\infty P(n,\tm) \Pi_n^2(\tm)\,    \Theta \big(\Pi_n(\tm)\big) \\
    &= \Delta E^2 \sum_{n=n'}^\infty \frac{\Big[P(1,n,\tm)-P(0,n,\tm)\Big]^2}{P(n,\tm)},
\end{align}
where we introduced $n'$ as the lowest value of $n$ for which population inversion occurs, i.e., for which the conditional polarisation $\Pi_n(\tm)$ is positive. In the following, we drop the explicit dependence on $\tm$ in the notations.

Given the work moments~\cref{eq:work_extracted_specific} and~\cref{ergotropysecond}, the noise-to-signal ratio is
\begin{align}
\label{noise}
   \frac{\Delta \Wext^2}{\langle \Wext\rangle ^2} = \frac{\langle \Wext^2 \rangle}{\langle \Wext \rangle ^2} - 1,
\end{align}

where $\Delta \Wext^2 \equiv \langle \Wext^2 \rangle - \langle \Wext\rangle ^2$ is the variance. The inverse of the noise-to-signal ratio is often referred to as the precision~\cite{SEIFERT2018,Berx_2024_2}. For measurement outcomes $n\geq n'$, the conditional probability of finding the \gls{tls} in the excited state exceeds that of the ground state, i.e., $P(1|n,\tm) > P(0|n,\tm)$, and the conditional polarisation becomes positive. This is a necessary condition for nonzero work extraction from the system. Averaging over all cycles then leads to an overall positive work extraction. Finally, the engine cycle is completed by resetting both the \gls{tls} and the \gls{qho} to their initial states by rethermalisation through interaction with their respective thermal reservoirs.

\section{\label{sec:results} Pareto optimisation of performance}

The \emph{average} work that can be extracted is bounded by the information gained in the measurement, $\langle\Wext\rangle\le\Temp[S]\,I$, as we have shown in Ref.~\cite{Hagman2025}. Extracting more work requires more information gained about the state of the system. However, since the measurement outcomes to gain this information are themselves stochastic, it is not guaranteed to extract the same amount of work in every cycle. It is therefore necessary to quantify how reliably the work extraction proceeds through measurement by means of the variance of the extracted work. Maximising the average work and minimising the work variance, i.e., the work fluctuations, are therefore genuinely competing objectives. The result is that a trade-off, i.e., a Pareto front between the squared average extracted work, $\langle\Wext\rangle^2$, and its fluctuations, $\Delta \Wext^2$, emerges, which we computed by using the NSGA-II genetic algorithm~\cite{Deb2008}. 
\begin{figure}[htp]
    \centering
    \includegraphics[width=\linewidth]{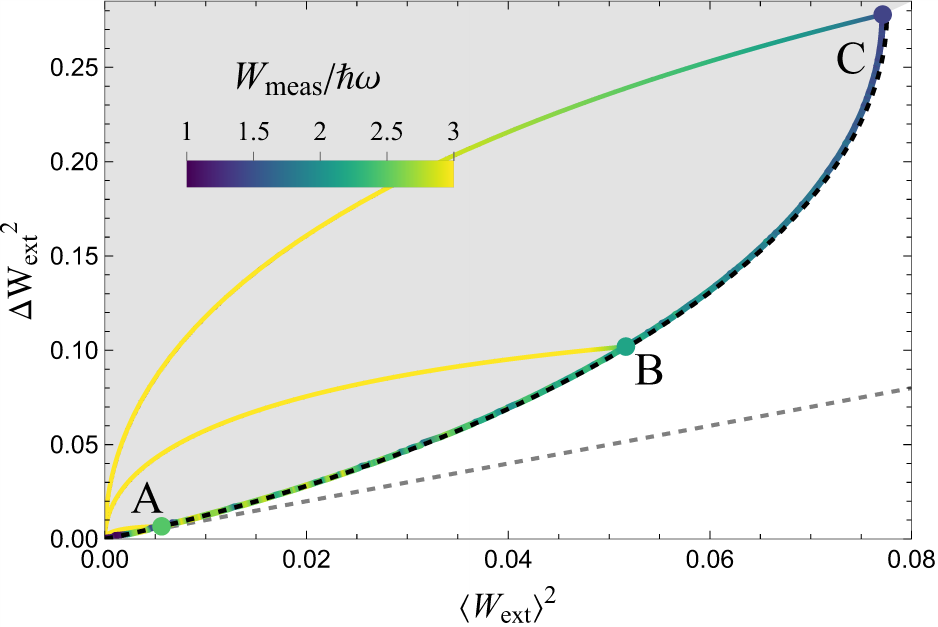}     
    \caption{Pareto front (thick, colour gradient curve, starting at the origin going through points $A$, $B$ and $C$) for simultaneously minimising $\Delta \Wext^2$ and maximising $\langle \Wext\rangle^2$ (measured in appropriate units of $\kb \Temp[S]$). All sub-optimal engine configurations lie above the front in the shaded region. Three sets of sub-optimal configurations obtained by changing the dimensionless measurement time $\omega \tm$ starting away from their Pareto-optimal points $A$, $B$ and $C$ are shown by full lines. The dashed gray line indicates $\Delta \Wext^2 / \langle \Wext\rangle ^2 = 1$; the dashed black line is the Pareto front corresponding to the exact limit $\lambda\to\infty$, Eq.~\eqref{eq:collapse2}. The colour scale shows that the measurement cost, $W_{\rm meas}$~\eqref{eq:measurement_work_specific}, decreases when allowing for large work fluctuations.}
    \label{fig:Pareto_work_fluctuations}
\end{figure}

\cref{fig:Pareto_work_fluctuations} portrays the resulting trade-off between $\Delta \Wext^2$ and $\langle \Wext \rangle^2$ resulting from the numerical optimisation protocol.
It can be seen that near the origin of the Pareto front its slope is close to unity (dashed gray line in~\cref{fig:Pareto_work_fluctuations}). This implies that the noise-to-signal ratio, i.e., the ratio of the variance to the mean squared, is at least $1$. Physically, this means the variance of the per-cycle work is no smaller than its mean squared: the device cannot act as a low-noise, steady work source. This result can be interpreted within our proposed information-engine cycle: the work $\Wext(\tm|n)$ is conditionally extracted depending on the measurement outcome $n$, such that nonzero work arises only from a subset of “successful” cycles. While such behaviour is undesirable for classical or quantum heat engines designed to deliver steady work output~\cite{Anka2024,Bouton2021}, the present information engine is more appropriately viewed as a stochastic rectifier of the thermal fluctuations that arise from the two-level system's contact with a single heat reservoir---a freely available thermodynamic resource~\cite{Chit2019}. In particular, this reservoir may originate from dissipation induced by another quantum process. The engine converts this otherwise unused energy into useful work whenever a favourable   measurement outcome occurs. Within this operational paradigm, a high degree of variability in $\Wext(\tm|n)$ is not only acceptable but intrinsic to the device’s function.

Moving away from the origin in~\cref{fig:Pareto_work_fluctuations}, the slope of the Pareto front increases monotonically, reflecting diminishing returns in extracted work: each additional unit of mean work entails a progressively larger increase in fluctuations. Notably, moving along the front towards \textit{lower} fluctuations comes with a \textit{higher} measurement cost, $W_{\rm meas}$~\eqref{eq:measurement_work_specific}. This highlights the energy price to pay for reduced fluctuations. 

\begin{figure}[htp]
    \centering
    \includegraphics[width=\linewidth]{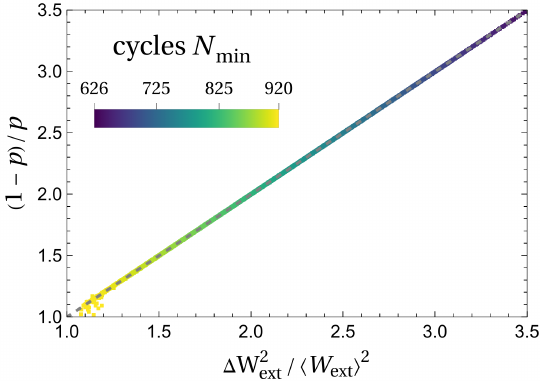}
    \caption{Collapse of the Pareto front (coloured points) according to equation~\eqref{eq:collapse} onto a linear function (dashed line), in the cold meter limit $\hbar\omega/\kb\Temp[M] \gg 1$. From the collapse, it can be seen that indeed the noise-to-signal ratio is bounded from below by unity. The colouring shows the minimal number of engine cycles $N_{\rm min}$ needed to resolve the corresponding operating point on the front with accuracy $\epsilon=0.05$ in the estimated success fraction $q$, Eq.~\eqref{eq:succes_fraction}, within failure tolerance $\delta = 0.01$ according to Eq.~\eqref{eq:cycle_nr}.}
    \label{fig:collapse}
\end{figure}

Interestingly, we found through our numerical optimisation algorithm that on the Pareto front the condition $\hbar\omega/\kb\Temp[M] \gg 1$ always holds. Physically, this corresponds to a meter at low temperatures, where we can assume that initially only the ground state of the oscillator is occupied, that is, $m=0$ and $P_{m=0}=1$. \cref{eq:probability0} and \cref{eq:probability1} can  consequently be written as
\begin{equation}
\label{eq:problow}
\begin{split}
    P(0,n,\tm) &=a \delta_{n,0}\,, \\ 
    P(1,n,\tm) &=b\left(\frac{1}{n!}\right)\lambda^{n}e^{-\lambda}\,,
\end{split}
\end{equation}
with $\lambda \equiv |\alpha|^2 =  \frac{g_{\rm eff}^2}{\hbar\omega}(1-\cos{(\omega \tm)})$ the effective measurement strength.
\cref{eq:work_extracted_specific} together with \cref{eq:problow} show that for cold meters one can always extract work $\Delta E$ for a measurement outcome $n\geq1$ which is obtained with probability $p \equiv b\left[1-e^{-\lambda}\right]$, while for $n=0$ (with probability $1-p$) no work is extracted. This can be translated into the Bernoulli work distribution
\begin{equation}
    \label{eq:work_distribution_cold}
    P(\Wext(\tm|n)) = 
    \begin{cases}
      p\,, & \Wext(\tm|n) = \Delta E \\
        1-p \,, & \Wext(\tm|n) = 0\,.
    \end{cases}
\end{equation}
From this distribution, computing the first two moments yields 
\begin{align}
    \langle \Wext\rangle &= \Delta Eb\bigg[1-e^{-\lambda} \bigg]\,,\label{eq:coldwork} \\ \langle \Wext^2\rangle &= \Delta E^2b\bigg[1-e^{-\lambda} \bigg] \label{eq:coldfluctuations},
\end{align}
for the average work and average squared work, respectively. 
The noise-to-signal ratio in this limit can therefore be computed as
\begin{align}
    \label{eq:collapse}
    \frac{\Delta \Wext^2}{\langle \Wext\rangle^2} = \frac{1- b \bigg[1-e^{-\lambda}\bigg]}{b \bigg[1-e^{-\lambda}\bigg]}=\frac{1- p}{p}.
\end{align}
Fig.~\ref{fig:collapse} confirms the equality of the numerically computed Pareto front data to the exact equation~\eqref{eq:collapse}. In the limit of $\lambda\to\infty$ we find a lower bound on the noise-to-signal ratio~\eqref{eq:collapse} by

\begin{align}
    \label{eq:collapse2}
    \frac{\Delta \Wext^2}{\langle \Wext\rangle^2} \geq \frac{1- b}{b}=e^{\frac{\Delta E}{k_{\rm B}T_{\rm S}}}.
\end{align}

This equation confirms the numerically found result that the noise-to-signal ratio is lower bounded by $1$, see Fig.~\ref{fig:Pareto_work_fluctuations}. Interestingly, Eq.~\eqref{eq:collapse2} reduces to the relative energy fluctuations of a qubit prepared in its initial thermal state, with probability $b$ to occupy the excited state and $a=1-b$ to occupy the ground state. In this limit, the meter is perfectly accurate and allows for extracting not only the average work but all higher moments as well from a single thermal bath. Equation~\eqref{eq:collapse2} therefore constitutes a \emph{thermodynamic} lower bound on the relative work fluctuations, determined entirely by the thermal excited-state population $b$ of the qubit. 

This shows that, even in the limit of perfect measurements, the fluctuations of the extracted work are not set by measurement imperfections but are instead fundamentally constrained by the same thermal resource from which the work is extracted. In particular, no improvement of the measurement can reduce the noise-to-signal ratio below $e^{\Delta E/\kb\Temp[S]}$. 

It is important to note, however, that within both our numerical optimisation framework and realistic experimental implementations, $\lambda$ can never truly diverge. Such a limit would require either an unbounded coupling strength or a vanishing energy-level spacing of the meter, neither of which is physically attainable. Consequently, we focus throughout on finite values of $\lambda$, while referring to the limit $\lambda\to\infty$ as an \emph{idealised} regime that nonetheless permits additional analytical insight. 

Thus, in this limit we can compute the resulting trade-off in a closed form, since it is effectively parametrised solely by $\Delta E/k_{\rm B}T_{\rm S}$. This trade-off is shown by the black dashed line in Fig.~\ref{fig:Pareto_work_fluctuations}. It constitutes the minimal idealised trade-off between work and work fluctuations; all realistic scenarios span Pareto fronts that lie above the $\lambda\to\infty$ result. 

\begin{figure}[htp]
    \centering
    \includegraphics[width=\linewidth]{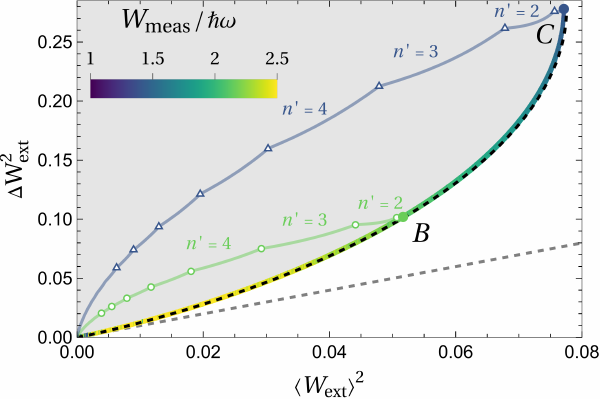}
    \caption{Pareto front (thick, colour gradient curve) for simultaneously minimising $\Delta \Wext^2$ and maximising $\langle \Wext\rangle^2$ (measured in appropriate units of $\kb \Temp[S]$). All sub-optimal engine configurations lie above the front in the shaded region; sets of sub-optimal configurations obtained by changing the ratio $\Temp[M]/\Temp[S]$ away from their Pareto-optimal values are shown by full lines. Engine designs where $n'$ changes along sub-optimal lines are indicated by open symbols, i.e., where the threshold $n'$ jumps to the next integer. The dashed gray line indicates $\Delta \Wext^2 / \langle \Wext\rangle ^2 = 1$; the dashed black line is the Pareto front corresponding to the exact limit $\lambda\to\infty$. The colour scale shows that the measurement cost, $W_{\rm meas}$~\eqref{eq:measurement_work_specific}, decreases when allowing for large work fluctuations.}
    \label{fig:suboptimal_x}
\end{figure}

\textbf{Variation from Pareto front---}
We next examine how engine designs depart from the Pareto-optimal front when a single parameter of the system-meter setup is varied. As shown by fixed-colour curves in~\cref{fig:Pareto_work_fluctuations}, adjusting, e.g., the measurement time moves the operating point off the front into the shaded sub-optimal region. Similar sub-optimal ``trajectories'' are shown for points $B,\,C$ in Fig.~\ref{fig:suboptimal_x} for varying the temperature ratio $\Temp[M]/\Temp[S]$, i.e., moving away from the optimal cold-meter limit. From the figure, we see that moving away from the Pareto-optimal cold meter limit by changing the temperature ratio, the population-inversion condition, $P(1|n,\tm) > P(0|n,\tm)$, the necessary condition for work extraction by ergotropy in Eq.~\eqref{eq:work_ext_TLS}, shifts to larger values of $n$. The threshold $n'$ marks the first value of $n$ at which this condition is satisfied. This behaviour was also observed in our earlier work~\cite{Hagman2025}. As a result, the trajectories develop convex branches in which $n'$ changes in discrete steps, starting from $n'=1$ at the Pareto front.

\textbf{Experimental realisability and finite-sample effects---}
While the Pareto front identified above characterises the optimal trade-off between the \emph{mean} extracted work and its variance, any experimental realisation necessarily operates over a finite number $N$ of cycles. For quantum engines, time is itself a resource~\cite{Taranto2023}, quantified operationally as the number of control steps; in a cyclic implementation, this corresponds directly to the number of engine cycles. The number of cycles $N$ therefore constitutes a resource, since the total experiment time $\mathrm{T}$ is realistically bounded from below by $\mathrm{T} \geq N \tm$. It is therefore essential to quantify how reliably this asymptotic Pareto trade-off can be inferred from a \emph{finite} sample.

We consider $N$ independent cycles, where each cycle yields either zero work or a work quantum $\Delta E$, such that $W_i = \Delta E\, \xi_i$, with $\xi_i \in \{0,1\}$ a Bernoulli random variable with success probability $p=b\left[1-e^{-\lambda}\right]$ as defined above. The number of successful cycles is $K=\sum_{i=1}^N \xi_i$, the total extracted work is $W_N = K\Delta E$, and the empirical success fraction reads
\begin{equation}
    \label{eq:succes_fraction}
    q \equiv \frac{K}{N} = \frac{W_N}{N\Delta E}.
\end{equation}
The finite-$N$ Pareto front is then obtained by replacing $p$ by its empirical estimate $q$ in Eqs.~\eqref{eq:coldwork} and~\eqref{eq:coldfluctuations}.

Since $q$ is the sample mean of independent and identically distributed (i.i.d.)\ Bernoulli trials, it obeys a large-deviation principle with probability distribution $P$ of the form $P(q \simeq x) \asymp \exp[-N \mathcal{J}(x)]$ (see Appendix~\ref{app:large_dev}). $P(q \simeq x)$ allows one to quantify the probability that an experiment using a finite number of cycles $N$ returns an \emph{apparent} operating point $x$. For a point $x\neq p$ we operate away from the true Pareto front, in the suboptimal regime. Imposing an absolute accuracy $\epsilon$ and failure tolerance $\delta$ via $P(|q-p|>\epsilon)\le \delta$ yields the estimate
\begin{equation}
    \label{eq:cycle_nr}
    N \gtrsim \frac{2p(1-p)}{\epsilon^2}\ln\!\frac{1}{\delta}.
\end{equation}

For fixed $(\epsilon,\delta)$, the required number of cycles is governed by the factor $2p(1-p)$. The prefactor $2p(1-p)$ reaches its maximum at $p=1/2$ at the origin of the Pareto front. Along the front the success probability decreases monotonically, from $p\to1/2$ at the origin to $p\approx0.218$ at the maximum-work point, so that $2p(1-p)$ falls monotonically as the work output grows. Consequently, achieving higher precision, i.e., smaller work fluctuations, demands a larger number of engine cycles to resolve a point on the front. This is shown in Fig.~\ref{fig:collapse}, where at the origin, i.e., for high precision, we find numerically that the minimum cycle number $N_0 = 920$, while in the maximum-work limit we find that the minimum cycle number $N^* = 626$.

In the idealised limit $\lambda \to \infty$, the probability $p$ becomes exactly equal to the excited-state population $b$, which is directly related to the qubit gap $\Delta E/\kb\Temp[S]$. At the Pareto-optimal point of maximal extracted work, which can be determined analytically by maximising Eq.~\eqref{eq:coldwork}, the resulting value of $b$ can be found and inserted into Eq.~\eqref{eq:cycle_nr}. One then finds 
\begin{equation}
    N^* \gtrsim 0.341 \, \frac{\ln(1/\delta)}{\epsilon^2}.
\end{equation}
Requiring, for instance, an accuracy $\epsilon = 0.05$ and failure tolerance $\delta = 0.01$ gives $N^* \gtrsim 628$. By contrast, at the origin of the Pareto front, where $p = 1/2$, one finds $N_0 \gtrsim 921$ for the same parameters. Comparing these numbers to the ones found numerically for the realisable Pareto front, we find them to be very close. The ratio of required cycle numbers can be computed as
\begin{equation}
\label{eq:overhead}
    \frac{N_0}{N^*} = \frac{1}{4b^*(1-b^*)} \approx 1.467,
\end{equation}
providing a parameter-free estimate of the overhead in resolving the low-work regime.

\section{Information Flow and Geometry}\label{sec:information}

{\bf Information flow along the Pareto front---} The information gained about the \gls{tls} by the \gls{qho} meter is given by the mutual information $I$, defined in Eq.~\eqref{eq:InformationGain}. It quantifies the correlations established between the system and meter during the coupling stage, and thus measures how much information about the system state is available for feedback. In the Pareto-optimal regime, attained in the cold-meter limit where the measurement becomes effectively binary, $I$ can be computed by inserting Eq.~\eqref{eq:problow} into Eq.~\eqref{eq:InformationGain}. The resulting information gain specified to the Pareto-optimal solutions $I_{\rm opt}$ then reads
\begin{align}
\label{eq:I_opt}
    I_{\rm opt} = -\kb \bigg[& b\ln{b}-be^{-\lambda}\ln{(be^{-\lambda})}\\ \notag &+(a+be^{-\lambda})\ln{(a+be^{-\lambda})}\bigg]\,,
\end{align}
where $\lambda =  \frac{g_{\rm eff}^2}{\hbar\omega}(1-\cos{(\omega \tm)})$.

\begin{figure}[htp]
    \centering
    \includegraphics[width=\linewidth]{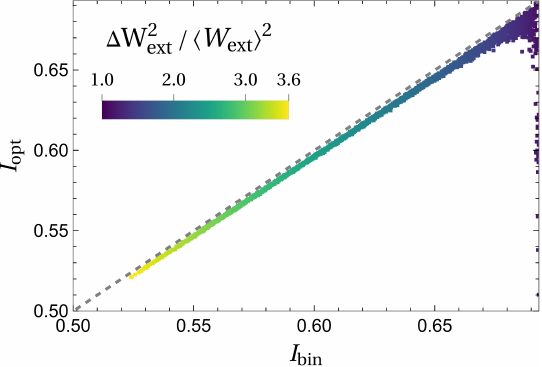}
    \caption{Mutual information along the Pareto-optimal front, $I_{\rm opt}$, plotted against the binary Shannon entropy $I_{\rm bin}$ (both in nats), exhibits a clear collapse onto a linear relation (gray, dashed line) for large $\langle\Wext\rangle$. The colour scale encodes the corresponding Pareto-optimal average extracted work, $\langle \Wext \rangle^2$, and indicates that the collapse becomes most pronounced in the high-work regime.}
    \label{fig:binary_entropy}
\end{figure}

First, we observe that achieving a smaller noise-to-signal ratio in the extracted work, $\Delta \Wext^2/\langle \Wext \rangle ^2 $, requires a larger information gain $I_{\rm opt}$, as illustrated in \cref{fig:binary_entropy}.

Moreover, we find that Eq.~\eqref{eq:I_opt} is upper bounded by 
\begin{align}
\label{eq:I_opt2}
   I_{\rm opt} \leq I_{\rm bin} = -\kb \left[p \ln{p} + (1-p)\ln{(1-p)}\right],
\end{align}
with the bound being saturated in the limit $\lambda\to\infty$, as shown in~\cref{fig:binary_entropy}. In this figure, the information gain is plotted as a function of the binary Shannon entropy $I_{\rm bin}$ for the points lying on the Pareto front shown in~\cref{fig:Pareto_work_fluctuations}.

{\bf Information geometry---} In contrast with the mutual information, the Fisher information $\mathcal{I}(\theta)$ is a measure of the local sensitivity of the entire information engine as a response to slight changes in one of the underlying parameters $\theta$. It determines how much the meter distribution changes when one perturbs a parameter $\theta$ by a small amount $\mathrm{d}\theta$. High values of $\mathcal{I}(\theta)$ thus indicate that the engine's statistics are highly sensitive to small changes in $\theta$. Let us for now set the measurement time $\tm$ as the parameter to be changed. The Fisher information then defines a distance metric on the smooth statistical manifold of distributions $P(n,\tm)$. The distance $\mathrm{d}s$ between two distributions $P(n,\tm)$ and $P(n,\tm+\mathrm{d}\tm)$ is then defined through $\mathrm{d}s^2 = \frac{1}{2}\mathcal{I}(\tm) \mathrm{d}\tm^2$, with $\mathcal{I}(\tm) \equiv \langle (d_{\tm}\ln{P(n,\tm))^2}\rangle $; changing $\tm$ thus corresponds to moving along a curve in this manifold, with $\mathcal{I}(\tm)$ specifying how fast distributions change.

The maximal speed by which one can effect such changes is given by $v_F := \sqrt{\mathcal{I}(\tm)/2}$. In the cold meter temperature limit, it can be computed exactly, see~\cref{eq:Fisher_cold_meter_app} in App.~\ref{app:Cold}, as 
\begin{equation}
    \label{eq:Fisher_cold_meter}
    \mathcal{I}(\tm) = b \dot{\lambda}^2 \left[\frac{1}{\lambda} - \frac{a e^{-\lambda}}{a+b e^{-\lambda}} \right]\,,
\end{equation}
where the overdot denotes differentiation with respect to $\tm$.

\begin{figure}[htp]
    \centering
    \includegraphics[width=\linewidth]{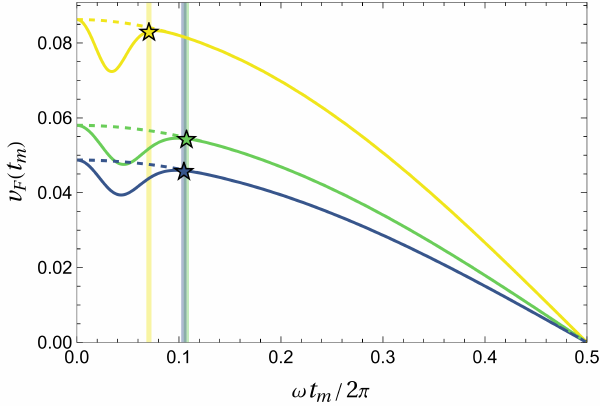}     
    \caption{The speed $v_F(\tm)$, measured in units of $k_B T_S/\hbar$, as a function of the unconstrained measurement time for the three selected points on the Pareto front of~\cref{fig:Pareto_work_fluctuations}: A (yellow line; low work), B (green line; intermediate work) and C (blue line; high work). Vertical lines indicate the corresponding optimal $\tm$ for those points on the Pareto front, with symbols showing the corresponding value of $v_F(\tm)$. Dashed lines correspond to an asymptotic approximation for $\lambda\gg1$. It can be seen that the Pareto-optimal points lie very close to the measurement time corresponding to the local maximum of the speed, i.e., the optimal engine designs lie very close to the speed limit.}
    \label{fig:Pareto_fisher}
\end{figure}

In Fig.~\ref{fig:Pareto_fisher} the speed $v_F(\tm)$ is shown for three sets of parameters obtained from the Pareto-optimal points labelled A, B and C in~\cref{fig:Pareto_work_fluctuations}, as a function of the measurement time $\tm$. The speed is a non-monotonic function of measurement time, exhibiting a local minimum followed by a local maximum at higher $\tm$. For $\tm = 0$, the speed is globally maximal, while for $\tm = \pi/\omega$ it becomes identically zero. Note that $v_F$ is symmetric around $\tm = \pi/\omega$. 

Interestingly, even though the Fisher information is not directly related to the work Pareto front, the Pareto-optimal points do appear close to local maxima of the speed, as shown by vertical lines in Fig.~\ref{fig:Pareto_fisher}. This signifies that optimal engine designs in the work vs. fluctuations sense are naturally close to the most time-sensitive operating points. 

In the limit $\lambda \gg 1$ we can expand the Fisher information as $\mathcal{I}(\tm) = \frac{g_{\rm eff}^2 \omega}{\hbar}b [1+\cos{(\omega \tm)}]$. The dashed lines in Fig.~\ref{fig:Pareto_fisher} show that this expansion is a monotonic function, contrary to the full expression~\eqref{eq:Fisher_cold_meter}. 
The local maximum observed in the full expression thus vanishes. This shows that the full nonlinear model~\eqref{eq:Fisher_cold_meter} can produce structure in the sensitivity landscape that an asymptotic approximation washes out. We therefore need to look beyond single-parameter sensitivity.

Interestingly, in the low-meter-temperature limit, the Fisher information $\mathcal{I}(\tm)$ can be used to establish an upper bound on the average extracted work $\langle \Wext(\tm) \rangle$ per measurement time $\tm$ (see Appendix~\ref{app:cramerRao} for details)
\begin{align}
\label{eq:CramerRaoPower}
 \frac{\langle \Wext(\tm) \rangle}{\tm}  \leq \sqrt{\langle \Delta W_{\rm ext}^2 \rangle_{\tm} \langle \mathcal{I}(\tm) \rangle_{\tm}}.
\end{align}
Here $\langle X \rangle_{\tm} :=\tm^{-1} \int_0^{\tm} X dt$  denotes the time average of the Fisher information and of the extracted-work fluctuations. Equation~\eqref{eq:CramerRaoPower} can be interpreted as an upper bound on the average extracted power~\cite{Hagman2025} from the engine. A finite extracted power therefore requires the product of the time-averaged work fluctuations and Fisher information to remain finite.

{\bf Control geometry and parameter sloppiness---}While the single-parameter sensitivity analysis with respect to the measurement time $\tm$ captures the intrinsic speed limit of the engine, we can extend the geometric picture to the full parameter space, with parameters collected into the vector $\bm{\theta}$. The infinitesimal statistical distance is defined by $\mathrm{d}s^2=\frac{1}{2}\mathcal{I}_{ij}(\bm{\theta})\,\mathrm{d}\theta^i \mathrm{d}\theta^j$, with
\begin{equation}
\mathcal{I}_{ij}(\bm{\theta})=-\big\langle \partial_{\theta_i}\partial_{\theta_j}\ln P(n,\tm)\big\rangle
\end{equation}
the Fisher information matrix (FIM), and where we used the Einstein summation convention. The FIM's spectral decomposition,
\begin{equation}
\mathcal{I}(\bm{\theta}) = V M V^\intercal,
\qquad
M=\diag(\mu_1\geq\dots\geq\mu_N\geq0),
\end{equation}
whose eigenvalues $\mu_i$ identify \emph{stiff directions} ($\mu_i\gg 0$), along which small parameter changes strongly affect the output statistics, and \emph{sloppy directions} ($\mu_i\approx 0$), where the response is weak.

In the cold-meter limit, the FIM can be computed analytically (see Appendix~\ref{app:FIM}) and has $\rank(\mathcal{I})\leq 2$. Thus, at most two directions are stiff and at least two are sloppy, since the model depends effectively only on the two dimensionless combinations $a=(1+e^{-\beta_\text{S}\Delta E})^{-1}$ and $\lambda = \frac{g_{\rm eff}^2}{\hbar\omega}(1-\cos{(\omega \tm)})$. The TLS gap $\Delta E/k_B T_S$ directly controls the occupation probability $a$, whereas $\hbar\omega$, $g_{\mathrm{eff}}$, and $\tm$ enter only through the single effective measurement strength $\lambda$. In other words, the apparent four-dimensional control space collapses onto a two-dimensional effective parameter space. This parametric reduction shows that all parameters of an optimally operating information engine can be compressed into a single parameter pertaining to the work extraction device, the qubit, and another pertaining to the measurement device, the harmonic oscillator, and the interaction.

The sloppy directions correspond to coordinated parameter changes that leave both $a$ and $\lambda$ unchanged. For instance, a shorter measurement time can be compensated by a stronger coupling, or a larger oscillator scale by a weaker coupling, without affecting the statistics. This degeneracy provides substantial experimental flexibility: since the coupling strength $\geff$ and frequency $\omega$ are typically determined by hardware constraints, sub-optimal hardware design can be compensated for by adjusting the measurement time $\tm$, effectively staying on the Pareto front despite these limitations.

In the strong-measurement limit $\lambda\to\infty$, sensitivity along the $\lambda$-direction also decreases, and the Fisher geometry becomes effectively one-dimensional; beyond a certain point, the measurement saturates and further increases in strength barely alter the statistics.

\section{\label{sec:conclusion}Conclusions \& Outlook}

We have analysed a finite-time quantum information engine in which work is extracted from a two-level system conditionally on the outcome of a measurement performed by a quantum harmonic oscillator meter. By combining multi-objective optimisation with an explicit analysis of the underlying measurement statistics, we identified the fundamental trade-off between average extracted work and work fluctuations. The resulting Pareto front shows that increasing the mean work output inevitably comes at the expense of larger fluctuations, while reducing fluctuations requires accepting a lower average output, which in turn has a larger measurement cost and requires a higher number of engine cycles. In this sense, the engine does not behave as a steady power source, but rather as an intermittent energy-to-work converter whose work output is fluctuating around a mean which is limited by information-theoretic constraints. These conclusions are particularly visible in the Pareto-optimal cold-meter regime, where the model becomes analytically tractable and the work statistics reduce to an effectively binary process.

The information analysis provides a complementary perspective on the same trade-off. Along the Pareto front, the mutual information between system and meter increases as the work fluctuations are reduced, showing that improved precision requires stronger information acquisition. At the same time, the Fisher-information analysis reveals that Pareto-optimal operating points for the engine lie close to regions of maximal sensitivity with respect to the measurement time, where small control changes have the strongest impact on the engine statistics. In the cold‑meter limit, the full control space collapses to an effective two‑dimensional description, making explicit which parameters, or, equivalently, which two independent parameter combinations, govern the engine’s performance. Within this regime, we identify an effective measurement strength that unifies the measurement time, the system–meter coupling, and the oscillator frequency. This formulation provides substantial experimental flexibility by allowing different parameters to be tuned interchangeably to remain close to Pareto‑optimal operation.

Several natural directions for future research emerge from this work. In particular, harnessing the measurement‑controlled precision of quantum processes offers a promising route toward the design of, e.g., high‑fidelity quantum clocks, provided that all relevant \textit{thermodynamic costs}, such as energy expenditure and the number of required operational cycles, are fully accounted for. More broadly, we expect our framework to serve as a valuable guide for experiments on information‑powered quantum devices, where precision, fluctuations, and information flow must be jointly optimised within fundamental thermodynamic constraints.

\begin{acknowledgments}

We thank Janine Splettstoesser, Didrik Palmqvist and Ludovico Tesser for carefully reading our manuscript and providing valuable feedback.
Funding from the European Research Council (ERC) under the European Union’s Horizon Europe research and innovation programme (101088169/NanoRecycle) (H.K.) is gratefully acknowledged. J.B. is supported by the Novo Nordisk Foundation with grant No. NNF18SA0035142.
\end{acknowledgments}
\section*{Code Availability}
The code used to generate the data for this article is available on Zenodo (DOI: 10.5281/zenodo.20095787) under a Creative Commons Attribution 4.0 International (CC BY 4.0) license.

\appendix

\section{Large-deviation analysis}
\label{app:large_dev}

The number of successful work extraction events in $N$ successive, independent engine cycles is binomially distributed. Each cycle is modelled as a Bernoulli trial with success probability $p$, so the total number of successful cycles, $K$, is the sum of $N$ independent binary outcomes. Since the quantity of interest is the success fraction, we write $q=K/N$ and, equivalently, substitute $K=qN$ into the Binomial distribution. This is the natural variable to consider as it measures the observed success rate over the full set of cycles, such that the binomial distribution reads
\begin{equation}
\label{eq:Bernoulli}
    P(q ) = p^{qN} (1-p)^{N(1-q)} \frac{N!}{(N-qN)!(qN)!}.
\end{equation}
To study the probability of observing a given value of $q$ for large $N$, we evaluate the binomial distribution, Eq.~\eqref{eq:Bernoulli}, in the asymptotic regime where $N$ is large. In this limit, the leading behaviour of factorials appearing in the binomial coefficient is well captured by Stirling's approximation $\ln{N!}\approx N\ln{N} - N$. We use it to isolate the binomial distribution's dominant exponential dependence on $N$. This makes it possible to identify a large-deviation form: the probability of observing a value $q$ decays exponentially with $N$, namely
\begin{equation}
    P(q) \asymp \exp\!\big[-N \mathcal{J}(q)\big],
\end{equation}
with rate function \cite{TOUCHETTE2009}
\begin{equation}
    \mathcal{J}(q)= q \ln \frac{q}{p} + (1-q)\ln \frac{1-q}{1-p}.
\end{equation}

To quantify finite-sample reliability, we consider the probability of deviations larger than a prescribed tolerance $\epsilon$, i.e.,
\begin{equation}
    P(|q-p|>\epsilon).
\end{equation}
To leading exponential order this behaves as
\begin{equation}
    P(|q-p|>\epsilon)\asymp \exp\!\big[-N \mathcal{J}^*(\epsilon)\big],
\end{equation}
where
\begin{equation}
    \mathcal{J}^*(\epsilon)=\inf_{|q-p|>\epsilon}\mathcal{J}(q).
\end{equation}
Imposing a tolerance through $P(|q-p|>\epsilon)\le \delta$ yields the bound
\begin{equation}
    N \gtrsim \frac{\ln(1/\delta)}{\mathcal{J}^*(\epsilon)}.
\end{equation}

Close to its minimum at $q=p$, the rate function admits the quadratic expansion
\begin{equation}
    \mathcal{J}(q)\approx \frac{(q-p)^2}{2p(1-p)},
\end{equation}
such that
\begin{equation}
    \mathcal{J}^*(\epsilon)\approx \frac{\epsilon^2}{2p(1-p)}.
\end{equation}
Substituting this into the bound above yields
\begin{equation}
    N \gtrsim \frac{2p(1-p)}{\epsilon^2}\ln\!\frac{1}{\delta}.
\end{equation}
\section{Derivation of Cramér-Rao bound}\label{app:cramerRao}

Consider the time-derivative of the average work extraction~\cref{eq:work_extracted_specific}

\begin{align}
\label{eq:fisher}
  d&_{\tm} \langle \Wext(\tm) \rangle = \sum_n
\big[ (d_{\tm} \Wext(\tm|n))P(n,\tm) \\ \notag &+ \Wext(\tm|n)(d_{\tm} P(n,\tm))\big] \\ \notag 
&= 
 \langle d_{\tm} \Wext(\tm)\rangle + \sum_n \Wext(\tm|n)(d_{\tm} P(n,\tm)).
\end{align}

The first term in~\cref{eq:fisher} can be cast into 
\begin{align}
\label{eq:WorkDeriv}
\langle d_{\tm} \Wext(\tm)\rangle =& 2\Delta E\sum_n \frac{P(0,n,\tm)(d_{\tm} P(1,n,\tm))}{P(1,n,\tm)+P(0,n,\tm)} \\ \notag &\times\Theta \big(\Pi_n(\tm)\big),
\end{align}
 with $\Pi_n(\tm) \equiv P(1|n,\tm) -P(0|n,\tm)$. 
We can rewrite~\cref{eq:fisher} further to 
\begin{align}
\label{eq:fisher2}
 d_{\tm}\langle \overline{\Wext}(\tm) \rangle &\equiv  d_{\tm} \langle \Wext(\tm) \rangle - \langle d_{\tm} \Wext(\tm)\rangle \\ \notag &= \sum_n \Wext(\tm|n)(d_{\tm} P(n,\tm))\\ \notag 
 &= \sum_n \big[\Wext(\tm|n)-\langle \Wext(\tm) \rangle\big] \times \\ \notag &(d_{\tm} \ln{P(n,\tm)})P(n,\tm),
\end{align}
where we used that the average of $d_{\tm} \ln{P(n,\tm)}$ vanishes due to conservation of probability. Taking the square and applying the Cauchy-Schwarz inequality, we obtain the Cram\'{e}r-Rao bound
\begin{align}
\label{eq:CramerRao}
 (d_{\tm}\langle \overline{\Wext}(\tm) \rangle)^2 \leq& \big\langle (\Wext(\tm)-\langle \Wext(\tm) \rangle)^2\big\rangle \times \\ 
 \notag &\langle (d_{\tm}\ln{P(n,\tm))^2}\rangle \\ \notag 
 &=\Delta W_{\rm ext}^2 \mathcal{I}(t_m)  ,
\end{align}
where $\mathcal{I}(t_m) \equiv \langle (d_{\tm}\ln{P(n,\tm))^2}\rangle $ is the Fisher information. 

In the limit of a cold meter $\hbar\omega\gg k_B \Temp[M]$ the term $\langle d_{\tm} \Wext(\tm)\rangle$ in Eq.~\eqref{eq:WorkDeriv} vanishes given Eq.~\eqref{eq:problow} in the main text. 

Eq.~\eqref{eq:CramerRao} thus reads 
\begin{align}
\label{eq:CramerRao2}
 (d_{\tm}\langle \Wext(\tm) \rangle )^2 \leq\Delta W_{\rm ext}^2 \mathcal{I}(t_m).
\end{align}

The bound found in Eq.~\eqref{eq:CramerRao2} can further be transformed into a time-averaged version~\cite{Farre2020}
\begin{align}
\label{eq:CramerRao3}
 \frac{\langle \Wext(\tm) \rangle}{\tm}  \leq \sqrt{\langle \Delta W_{\rm ext}^2 \rangle_{\tm} \langle \mathcal{I}(t_m) \rangle_{\tm}}, 
\end{align}
where $\langle \Wext(\tm) \rangle$ is the cycle-averaged ergotropy extracted during the measurement time $\tm$, and
\begin{align}
    \langle X \rangle_{\tm} :=\tm^{-1} \int_0^{\tm} X dt. 
\end{align}
In fact the l.h.s of Eq.~\eqref{eq:CramerRao3} coincides with the average extracted power per measurement cycle in accordance with previous work \cite{Hagman2025}.

\section{Cold meter limit}
\label{app:Cold}

In the limit of a cold meter $\hbar\omega\gg k_{\rm B}T_{\rm M}$ the first term in Eq.~\eqref{eq:fisher} captured by Eq.~\eqref{eq:WorkDeriv} vanishes given the condition  Eq.~\eqref{eq:problow} in Sec.~\ref{sec:method_example}. The second term in Eq.~\eqref{eq:fisher} in this limit reads 
\begin{align}
\label{eq:WorkDeriv_app}
\Delta E&\sum_n \Pi_n(\tm)\Theta \big(\Pi_n(\tm)\big) d_{\tm} P(1,n,\tm) \\ \notag
&= \Delta E\sum_n d_{\tm} P(1,n,\tm)=\Delta Eb \dot{\lambda}(\tm)e^{-\lambda(\tm)},
\end{align}
with $\lambda =  \frac{g_{\rm eff}^2}{\hbar\omega}(1-\cos{(\omega t_m)})$ and $\Pi_n(\tm) = P(1|n,\tm) -P(0|n,\tm)$, so that we can rewrite~\cref{eq:fisher2} as
\begin{align}
 d_{\tm}\langle \overline{\Wext}(\tm) \rangle = \Delta Eb \dot{\lambda}(\tm)e^{-\lambda(\tm)}.
\end{align}
The fluctuations $\Delta \Wext^2 $ in the cold temperature limit reads
\begin{align}
    \label{eq:collapse_2}
\Delta \Wext^2 =& \Delta E^2 b \bigg[1-e^{-\lambda(t_m)}\bigg]\bigg[a+be^{-\lambda(t_m)}\bigg].
\end{align}
The Fisher information in the cold meter limit reads
\begin{align}
    \label{eq:Fisher_cold_meter_app}
    &\mathcal{I}(t_m) = \sum_{n=0}^\infty 
    \frac{(d_{\tm} P(1,n,\tm))^2}{P(0,n,\tm)+P(1,n,\tm)}\\ \notag 
    &=\dot{\lambda}(\tm)^2 \left[\frac{b^2 e^{-2\lambda(\tm)}}{a + b e^{-\lambda(\tm)}} + b \left(\frac{1}{\lambda(\tm)} - e^{-\lambda(\tm)}\right)\right]
    \\ \notag &=b \dot{\lambda}(\tm)^2 \left[\frac{1}{\lambda(\tm)} - \frac{a e^{-\lambda(\tm)}}{a+b e^{-\lambda(\tm)}} \right].
\end{align}

\section{Fisher information-based sloppiness analysis}\label{app:FIM}

By explicitly writing down the components of the FIM, $\mathcal{I}_{ij}(\bm{\theta})= -\langle\partial_{\theta_i}\partial_{\theta_j}\ln{P(n,\tm)}\rangle$, it can be seen that it admits the form $\mathcal{I}(\bm{\theta}) = J(\bm{\theta})^\intercal\mathcal{I}^{\rm red}(\bm{\theta})J(\bm{\theta})$, where $J$ is the Jacobian matrix of transformation of $(a,\lambda)$ with respect to the dimensionless parameters $(C_S,C_M,C_g^2,\tau)$, where $C_S = \Delta E/\kb\Temp[S]$, $C_M = \hbar\omega/\kb\Temp[S]$, $C_g^2 = \geff^2/\kb\Temp[S]$ and $\tau = \omega \tm/2\pi$. For notational simplicity, let us set $r= \exp{(-\lambda)}$. Now, $\mathcal{I}_F^{\rm red}$ is the reduced FIM with respect to $(a,\lambda)$ and can be explicitly computed as 
\begin{equation}
    \mathcal{I}^{\rm red} = \begin{pmatrix}
         \mathcal{I}_{aa} &  \mathcal{I}_{a\lambda} \\ 
          \mathcal{I}_{\lambda a} &  \mathcal{I}_{\lambda\lambda}
    \end{pmatrix}\,,
\end{equation}
where 
\begin{align}
    \mathcal{I}_{aa} &= \frac{1-r}{(1-a)(a + r (1-a))}\,,\\
    \mathcal{I}_{a\lambda} &= \mathcal{I}_{\lambda a} = -\frac{r}{a + r (1-a)} \\
    \mathcal{I}_{\lambda\lambda}&= -\frac{(1-a) (a + r (1-a + a\ln{r}))}{(a + (1-a) r) \ln{r}}\,.
\end{align}

In the limit where $\lambda\gg1$, the two eigenvalues of the reduced FIM scale as $\mu_1 \sim 1/a(1-a)$ and $\mu_2\sim (1-a)/\lambda$; for $\lambda\to\infty$ another direction thus becomes sloppy.


\clearpage

\bibliography{sources}

\end{document}